\documentclass[12pt,preprint]{aastex}
\usepackage{epsfig}

\shorttitle{Radiative effects on the first protogalaxies}
\shortauthors{Johnson, Greif, \& Bromm}

\begin{document}

\title{Local radiative feedback in the formation of the first protogalaxies}

\author{Jarrett L. Johnson\altaffilmark{1}, Thomas H. Greif\altaffilmark{1,2,3}, and Volker Bromm\altaffilmark{1}}

\altaffiltext{1}{Department of Astronomy, University of Texas at Austin, 2511 Speedway, Austin, TX 78712; jljohnson@astro.as.utexas.edu, vbromm@astro.as.utexas.edu}

\altaffiltext{2}{Institut f\"{u}r Theoretische Astrophysik, Universit\"{a}t Heidelberg, Albert-Ueberle-Strasse~2, 69120 Heidelberg, Germany; tgreif@ita.uni-heidelberg.de}

\altaffiltext{3}{Fellow of the International Max Planck Research School for Astronomy and Cosmic Physics at the University of Heidelberg}

\begin{abstract}
The formation of the first galaxies is influenced by the radiative feedback from the first generations of stars.  This feedback is manifested by the heating and ionization of the gas which lies within the H~II regions surrounding the first stars, as well as by the photodissociation of hydrogen molecules within the larger Lyman-Werner (LW) bubbles that surround these sources. Using a ray-tracing method in three-dimensional cosmological simulations, we self-consistently track the formation of, and radiative feedback from, individual stars in the course of the formation of a protogalaxy. We compute in detail the  H~II regions of each of these sources, as well as the regions affected by their molecule-dissociating radiation.  We follow the thermal, chemical, and dynamical evolution of the primordial gas as it becomes incorporated into the protogalaxy. While the IGM is, in general, optically thick to LW photons only over physical distances of $\ga$ 30 kpc at redshifts $z$ $\la$ 20, the high molecule fraction that is built up in relic H~II regions and their increasing volume-filling factor renders even the local IGM optically thick to LW photons over physical distances of the order of a few kiloparsecs. We find that efficient accretion onto Pop~III relic black holes may occur after $\sim$ 60 Myr from the time of their formation, by which time the photo-heated relic H~II region gas can cool and re-collapse into the 10$^6$ {\rm M}$_{\odot}$ minihalo which hosts the black hole.  Also, Pop~II.5 stars, postulated to have masses of the order of 10 {\rm M}$_{\odot}$, can likely form from this re-collapsing relic H~II region gas, but their formation may be suppressed by LW feedback from neighboring star-forming regions. Overall, we find that the local radiative feedback from the first generations of stars suppresses the star formation rate by only a factor of, at most, a few.
\end{abstract}

\keywords{stars: formation --- molecular processes --- H~II regions --- galaxies: formation --- early universe --- cosmology: theory}

\section{Introduction}
The formation of the earliest galaxies plays a key role in a number of the most important questions being addressed in cosmology today. 
The first galaxies are predicted to have been the dominant sources of the radiation which reionized the universe (e.g. Ciardi et al. 2006), 
and they may have hosted the majority of primordial star formation (Greif \& Bromm 2006; but see also Jimenez \& Haiman 2006).  They are the 
likely sites for the formation of the 
most metal-poor stars that have recently been found in our Galaxy (e.g. Christlieb et al. 2002; Beers \& Christlieb 2005; Frebel et al. 2005), and possibly for the first efficient 
accretion onto the stellar black 
holes (see Johnson \& Bromm 2007) which may have been the seeds for the $\sim$ 10$^9$ {\rm M}$_{\odot}$ black holes that are inferred at redshifts 
$z$ $\ga$ 6 (Fan et al. 2004, 2006). Furthermore, an understanding of the formation of the first 
galaxies is crucial for the interpretation of galaxies now beginning to be observed at $z\ga$ 6 (e.g. Mobasher et al. 2005; Iye et al. 2006; 
Bouwens \& Illingworth 2006), as well as of the 
objects at redshifts $z\ga$ 10 which are expected to be detected with upcoming 
telescopes such as the {\it James Webb Space Telescope} ($JWST$) (Gardner et al. 2006).  Among these 
systems there promise to be some of the first metal-free objects that will be observable, and as such it is important that theoretical predictions of 
their properties are made.  

What were the effects of the radiative feedback from the first generations of stars on the formation of the first 
galaxies?  It is now widely held that the first stars 
(termed Population~III) were likely very massive, and therefore emitted copious 
amounts of radiation which profoundly affected their surroundings (Bromm et al. 1999, 2002; Abel et al. 
2002; Yoshida et al. 2006; Gao et al. 2007).  Recent work has demonstrated that the H~II regions surrounding the first stars were able
to evacuate the primordial gas from the minihalos that hosted these objects
(Whalen et al. 2004; Kitayama et al. 2004; Alvarez et al. 2006; Abel et al. 2006).  
The impact of these H~II regions on second generation star formation is complex (e.g. Ricotti, Gnedin \& Shull 
2001; Oh \& Haiman 2003; Ahn \& Shapiro 2006; Susa \& Umemura 2006). While initially the density in these regions is suppressed and 
the gas within heated to $\ga 10^4$~K, vigorous molecule formation can take place once the gas begins to cool and recombine after the central Pop~III star has collapsed to fom a massive black hole, leading to the 
possibility of the formation of low-mass primordial stars (Nagakura \& Omukai 2005; O'Shea et al. 2005; Johnson \& Bromm 2006, 2007; 
Yoshida et al. 2007).

An additional radiative feedback effect from the first stars is the photo-dissociation of the fragile hydrogen molecules 
which allow the primordial gas to cool and collapse into minihalos, with virial temperatures $\la$ 8,000 K (e.g. Barkana \& Loeb 2001).  
 The effects of the molecule-dissociating radiation from the first 
stars can reach far beyond their H~II regions (e.g. Ciardi et al. 2000), 
and thus star formation in distant minihalos may have been delayed or quenched 
altogether (e.g. Haiman et al. 1997, 2000; Mackey et al. 2003).  Interestingly, however, while the general intergalactic medium (IGM) 
at the epoch of the first stars becomes optically thick to Lyman-Werner (LW) photons only over vast distances (e.g. Haiman et al. 2000; see also Glover \& Brand 2001), the 
high molecule 
fraction that persists inside the first relic H~II regions leads to a high optical depth to these photons, potentially allowing star 
formation to take place in minihalos down to lower redshifts than would otherwise be possible (Ricotti et al. 2001; Oh \& Haiman 2002; Machacek 
et al. 2001, 2003; Johnson \& Bromm 2007).   

In the present work, we self-consistently track the formation of, and the radiative feedback from, individual Pop~III stars in the course of the formation 
of a primordial protogalaxy.  
We compute in detail the H~II regions and LW bubbles of each of these sources, and follow the evolution of 
the primordial gas as it becomes incorporated into the protogalaxy.  In \S~2, we describe our numerical methodology.  Our results 
are presented in \S~3, while we summarize our conclusions and discuss their implications in \S~4.

\section{Methodology}

\subsection{Cosmological Initial Conditions and Resolution}
We employ the parallel version of 
GADGET for our three-dimensional numerical simulations.  This code includes a tree, hierarchical gravity 
solver combined 
with the smoothed particle hydrodynamics (SPH) method for tracking the evolution of the gas (Springel, Yoshida \& White 2001).  Along with H$_2$, 
H$_2^{+}$, H, H$^-$, H$^+$, e$^-$, He, He$^{+}$, and He$^{++}$, we have included the five deuterium species D, D$^+$, D$^-$, HD and HD$^-$, using the same 
chemical network as in Johnson \& Bromm (2006, 2007).     

We carry out a three-dimensional cosmological simulation of high-$z$ structure formation which evolves 
both the dark matter and baryonic components, initialized according to the $\Lambda$CDM model at $z$ = 100. 
As in earlier work (Bromm et al. 2003; Johnson \& 
Bromm 2007), we adopt the cosmological parameters $\Omega_{m}=1 - \Omega_{\Lambda}=0.3$, 
$\Omega_{B}=0.045$, $h=0.7$, and $\sigma_{8}=0.9$, close to the values 
measured by {\it WMAP} in its first year (Spergel et al. 2003). Here we use a periodic box with a comoving size $L$ = 460 $h^{-1}$ kpc, but unless stated explicitly, we will always refer to physical distances in the present work. Our simulation uses a number of particles $N_{\rm DM}$ = $N_{\rm SPH}$ = 128$^3$, where the SPH particle mass is $m_{\rm SPH}$ $\sim$ 740 ${\rm M}_{\odot}$.

We have determined the maximum density of gas that can be reliably resolved in this simulation by carrying out a cosmological simulation from $z$ = 
100, in which we allow the gas to cool and collapse into minihalos without including radiative effects.  We then compare the minimum resolved mass, 
which we take to be $\sim$ 64 $m_{\rm SPH}$, with the Bonnor-Ebert mass, given by (see, e.g., Palla 2002)      

\begin{equation}
M_{\rm BE}\simeq 700 {\rm M}_{\odot} \left(\frac{T}{200 {\rm \,K}}
\right)^{3/2}\left(\frac{n}{10^{4}{\rm cm}^{-3}}\right)^{-1/2} \mbox{\ ,}
\end{equation}
where $n$ and $T$ are the number density and temperature of the gas, respectively.   
As shown in Figure~1, the gas evolves according to the canonical behavior of primordial gas collapsing in minihalos (see, e.g., Bromm et al. 2002).  We expect the gas 
in our simulations with radiative feedback to behave similarly as it collapses to high densities, since it is the formation of, and cooling by, molecules 
which will drive the collapse in both cases. Thus, we take the maximum density that we can reliably resolve to be that at which the Bonnor-Ebert mass 
becomes equal to the resolution mass. As is evident in Figure~1, this criterion results in a maximum resolvable density of $n_{\rm res}$ $\sim$ 20 cm$^{-3}$.  This density is four orders of 
magnitude higher than the background mean density at a redshift of $z$ $\ga$ 15, and such overdensities only occur in the minihalos within which the first 
Pop III stars form (see, e.g., Bromm \& Larson 2004; Yoshida et al. 2006).  We take it here that one Pop~III star, assumed to have a mass of 100 {\rm M}$_{\odot}$, 
will form from this dense, collapsing primordial gas inside a minihalo, consistent with recent work which shows that, in general, only single stars are expected to form in minihalos under these conditions (Yoshida et al. 2006). 
Pop~III stars with this mass are predicted to directly collapse to a black hole, and therefore produce 
no supernova explosion (e.g. Heger et al. 2003), which allows us to self-consistently neglect the possibility of ejection of metals into the primordial gas.  We will consider this 
possibility in future work. In the present work, we focus on the radiative feedback from the first stars.

\subsection{Radiative Feedback}
In our simulations with radiative feedback, 
we assume that stars are formed in minihalos which acquire densities higher than $n_{\rm res}$ = 20 cm$^{-3}$. In order to account for the radiative feedback from a star formed in a minihalo, the gas surrounding the star is first photo-heated. We then calculate the extent of the H~II region, as well as of the LW bubble around the star. 
We carry out this procedure every time a star forms in the simulation. The Pop~III star will soon die, and we then let the simulation evolve once more, allowing recombination to take place in the relic H~II region, and for molecules to reform within the relic LW bubble.  We expect that this procedure will provide reliable results, as the $\la 3$~Myr lifetime of a Pop III star is short compared to the typical dynamical times of the gas in this simulation.

\subsubsection{Photoionization}
To account for the presence of a 100 {\rm M}$_{\odot}$ Pop~III star in the minihalos in which the gas collapses to a density of $n_{\rm res}$, we first 
photoheat and photoionize the gas within 500~pc of the gas particle which first reaches this density 
for a duration of the lifetime of the star, using the same heating and ionization 
rates as in Johnson \& Bromm (2007). Our choice of a 500~pc radius ensures that the entire gas within the source minihalo, with virial radius $\sim 150$~pc, is photoheated, but that we do not photoheat the dense, neutral gas in neighboring halos. Just as in this previous work, we reproduce the basic density and velocity structure of the gas 
within 500~pc of the central source that has been found in detailed one-dimensional radiation hydrodynamics calculations (Kitayama et al. 2004; Whalen et al. 2004).

Once this density structure is in place around the point source, we employ a ray-tracing technique to solve for the H II region that surrounds the star at the end of its life.  We cast rays in $N_{\rm ray}$ $\sim$ 100,000 directions from the central source, and divide each ray into 500 
segments.  Then, we add up all of the recombinations that take place over the course of the star's 
lifetime in each bin along each of 
the $N_{\rm ray}$ rays, taking the number of recombinations to be 

\begin{equation}
N_{\rm rec} = \alpha_{\rm B} n_{\rm mean}^2 \frac{4\pi}{N_{\rm ray}} t_{\rm *} r^2 dr  \mbox{\ ,}
\end{equation}
where $\alpha_{\rm B}$ is the case B recombination coefficient, $r$ is the distance of the bin from the star, $dr$ is length of the bin in the 
direction radial to the star, and $t_{\rm *}$ is the lifetime of the star, here taken to be 3 Myr (Schaerer 2002). We compute $n_{\rm mean}$, 
the average number density of hydrogen atoms in a bin, as 

\begin{equation}
n_{\rm mean} = \frac{\sum n_{\rm H}}{N_{\rm part}} \mbox{\ .}
\end{equation}
Here $N_{\rm part}$ is the number of SPH particles in the bin and $n_{\rm H}$ is the number density of hydrogen of the individual SPH particles in that bin.    

Next, we assume that the star radiates an equal number of photons in every direction, and we take the total number of ionizing photons that it radiates in its 
lifetime to be 

\begin{equation}
N_{\rm ion} = Q_{\rm ion} t_{\rm *}\mbox{\ ,}
\end{equation}
where we have chosen $Q_{\rm ion}$, the average number of ionizing photons emitted per second by the star, to be 10$^{50}$ s$^{-1}$ (see Bromm et al. 2001; Schaerer 2002).  We then add up the 
recombinations in all of the bins, along 
each of the rays, beginning with those closest to the star and moving outward, until the number of recombinations along a ray equals the 
number of ionizing photons that are emitted along that ray.  If the number of recombinations in the bin falls below the number of atoms in the bin, 
then we count the number of atoms in the bin against the number of photons as well.  Doing this for each of the rays, we solve in detail for the H~II region of the star.  We set the free electron fraction to unity for each of the SPH particles that lie within the H~II region.  We set the temperature of the 
SPH particles within the H~II region, but outside of the 500 pc photo-heated region, to $T$ = 18,000 K, roughly the value at the outer edge 
of the photo-heated region. As well, the fraction of molecules in the H~II region is set to zero, as we assume that all molecules are 
collisionally dissociated at the high temperatures in the H~II region.

\subsubsection{Photodissociation}
To find the region in which the LW radiation from the star destroys H$_2$ and HD molecules, the ``LW bubble'' in our terminology, we carry out a 
ray-tracing procedure similar to the one used to solve for the H~II region. We use the same bins as in that procedure, but 
now we evaluate the formation time of H$_{\rm 2}$ molecules in each bin and compare this both to the lifetime of the star and to the dissociation time of the molecules.  For each bin, we compute the H$_{\rm 2}$ formation time as 
\begin{equation}
t_{\rm form, H_2} = \sum \frac{n_{\rm {\rm H_2}}}{ n_{\rm H} (k_1 n_{\rm H_2^{+}} + k_2 n_{\rm H^-})}/N_{\rm part}\mbox{\ ,}
\end{equation}
where $n_{\rm H_2^{+}}$ and $n_{\rm H^-}$ are the number densities of H$_2^{+}$ and H$^{-}$, respectively. The sum is over all the particles in the bin, $N_{\rm part}$, and  
$k_1$ and $k_2$ are the rate coefficients for the following two main reactions that produce H$_2$:

\begin{displaymath}
{\rm H} + {\rm H_2^{+}} \to {\rm H^{+}} + {\rm H_2} \mbox{\ ,}
\end{displaymath} 

\begin{displaymath}
{\rm H} + {\rm H^{-}} \to {\rm e^-} + {\rm H_2} \mbox{\ .}
\end{displaymath} 
We adopt the following values for these rate coefficients (de Jong 1972; Karpas et al. 1979; Haiman et al. 1996): 

\begin{equation}
k_1 = 6.4 \times 10^{-10}   {\rm cm}^{3} {\rm s}^{-1} \mbox{\ ,}
\end{equation}

\begin{equation}
k_2 = 1.3 \times 10^{-9} {\rm cm}^{3} {\rm s}^{-1} \mbox{\ .}
\end{equation}

The dissociation time for the molecules is obtained by finding the flux of LW photons from a 100 {\rm M}$_{\odot}$ Pop III star, assumed to be a blackbody emitter with radius $R_{\rm *}\simeq 3.9 R_{\odot}$ and effective temperature $T_{\rm *}\simeq 10^5$~K (e.g., Bromm et al. 2001). The dissociation time for unshielded molecules at a distance $R$ from the star is then given by (Abel et al. 1997)

\begin{equation}
t_{\rm diss, H_2} \sim 10^5{\rm \,yr} \left(\frac{R}{{\rm 1 kpc}}\right)^2  \mbox{\ .}
\end{equation}

Next, we note that for molecules to be effectively dissociated by the LW radiation, the dissociation time of the molecules must be shorter than both the lifetime of the star and the formation time of the molecules.  Therefore, we compare all of these timescales for each bin along each ray and set the fraction of molecules to zero if $t_{\rm diss, H_2}$ $\la$ $t_{\rm form, H_2}$ and $t_{\rm diss, H_2}$ $\la$ $t_{\rm *}$.  If this condition is not satisfied, then the molecule fraction is left unchanged from its value before the formation of the star.  This allows for the possibility of the effective shielding of H$_2$ molecules because it accounts for the build-up of H$_2$ column density, for instance, in relic H~II regions or in collapsing minihalos where the formation time of H$_2$ is relatively short.

We take into account the effects of self-shielding by adding up the H$_{\rm 2}$ column density $N$(H$_{\rm 2}$) along the ray contributed by each bin in which the molecules are not effectively dissociated.  We then adjust the dissociation time for the molecules in shielded bins according to (Draine \& Bertoldi 1996):

\begin{equation}
t_{\rm diss, H_2} \sim 10^5{\rm \,yr} \left(\frac{R}{{\rm 1 kpc}}\right)^2  \left(\frac{N({\rm H_ 2})}{10^{14} {\rm \,cm^{-2}}}\right)^{0.75} \mbox{\ ,}
\end{equation}
when the column density of molecules between the bin and the star is $N$(H$_{\rm 2}$) $\ga$ 10$^{14}$ cm$^{-2}$. 

The ionic species H$^{-}$ and H$_2^{+}$, which are reactants in the main reactions which form H$_2$, can also, in principle, be destroyed 
by the radiation from the star.  The photo-dissociation times for these species are given in terms of the temperature of the star $T_{*}$, 
the source of thermal radiation in our case, and the distance from the star $R$, as (Dunn 1968; de Jong 1972; Galli \& Palla 1998)

\begin{equation}
t_{\rm diss, H_2^{+}} = 5 \times 10^{-2} T_{\rm *}^{-1.59} {\rm exp}\left(\frac{82000}{T_{*}}\right) \left(\frac{R}{R_{*}}\right)^{2} {\rm \ s} \mbox{\ ,}
\end{equation}

\begin{equation}
t_{\rm diss, H^{-}} = 9.1 T_{\rm *}^{-2.13} {\rm exp}\left(\frac{8823}{T_{*}}\right) \left(\frac{R}{R_{*}}\right)^{2} {\rm \ s} \mbox{\ .}
\end{equation}
For the $100 {\rm M}_{\odot}$ star, we find that $t_{\rm diss, H_2^{+}}$ $\sim$ 5 $\times$ 10$^3$ yr 
$(\frac{R}{1 {\rm kpc}})^{2}$  and $t_{\rm diss, H^{-}}$ $\sim$ 9 $\times$ 10$^2$ yr $(\frac{R}{{\rm 1 kpc}})^{2}$.  The formation 
times for these species, on the other hand, are

\begin{equation}
t_{\rm form, H^{-}} = n_{\rm H^{-}} \left(\frac{dn_{\rm H^{-}}}{dt}\right)^{-1} \sim 3 \times 10^3    {\rm yr} \mbox{\ ,}
\end{equation}

\begin{equation}
t_{\rm form, H_2^{+}} = n_{\rm H_2^{+}} \left(\frac{dn_{\rm H_2^{+}}}{dt}\right)^{-1} \sim 4 \times 10^3    {\rm yr} \mbox{\ ,}
\end{equation}  
for primordial gas at a temperature of $T$ = 100 K and a density of $n_{\rm H}$ = 10$^{-2}$ cm$^{-3}$, typical for gas at the outskirts of 
a collapsing minihalo.  These formation timescales become much shorter for gas deeper inside minihalos, where the densities and temperatures are 
generally higher.  It is in these regions, in and around minihalos, where the presence of molecules is most important for cooling the gas.  Within these regions the 
photo-dissociation times for these ionic species are less than their formation times only if they are located $\la$ 2 kpc from the star.  Thus, photo-dissociation of these species will become ineffective at distances $\ga$ 2 kpc from the star, a distance comparable to the size of the H~II region of a Pop III star (Alvarez et al. 2006; Abel et al. 2006).  Since we assume that molecules are 
collisionally destroyed inside the H~II region, and 
since the LW bubble will generally be larger than the H~II region, we ignore the photo-dissociation of H$^{-}$ and H$_2^{+}$ 
in our calculations.   
 
LW photons can also be absorbed by hydrogen atoms, through the Lyman series transitions, as discussed in detail by Haiman et al. (1997, 2000).  
However, this atomic absorption will only have a significant effect on the LW flux over distances large enough that the Hubble expansion 
causes many of the LW photons to redshift to wavelengths of the Lyman series transitions.  The light-crossing time for our cosmological box is much 
shorter than the Hubble time at the redshifts that we consider.  Thus, LW photons will be negligibly redshifted as they cross our cosmological box 
and we can safely neglect the minimal atomic absorption of these photons that may take place.  

It has also been found that a shell of H$_2$ molecules may form ahead of the expanding H~II regions surrounding the first stars (see Ricotti et al. 2001).  
These authors find that such shells may become optically thick to LW photons.  However, Kitayama et al. (2004) have 
discussed that such shells are 
likely short-lived, persisting for only a small fraction of the lifetime of the star. Thus, for our calculations we neglect the possible formation 
of such a shell, as we expect that the opacity to LW photons through this shell will be very small when averaged over the lifetime of the star.    
Additionally, as we show in section 3.1, the regions affected by the LW feedback from a single Pop III star extend, at most, only a few kiloparsecs beyond the H~II region of such a star, which itself extends $\sim$ 5 kpc.  If an H$_2$ shell forms ahead of the H~II region, then the extent of the LW bubble will only be suppressed by, at most, a factor of a few in radius. 

\subsection{Sink Particle Formation}
We have carried out two simulations, 
one with radiative feedback and one in which the simulation evolves without including radiative effects.  
For the former simulation, we allow stars to form when the density reaches $n_{\rm res}$, and 
the expansion of the gas around the star due to photo-heating suppresses the density so that our resolution limit is not violated.  For the simulation without 
radiative feedback we allow sink particles to form when the density of the gas reaches $n_{\rm res}$.  
Since the sink particles will 
form only in minihalos which are expected to form Pop III stars, we are able to track the star formation rate 
in the case without feedback by tracking the formation of 
sink particles.  We can then compare the sites, and rates, 
of star formation in each of the simulations in order to elucidate the effect that radiative feedback has on Pop III star formation.
 
\section{Results}
In this section we discuss the evolution of the primordial gas under the influence of the radiative feedback which arises as the first stars are 
formed in a region destined to subsequently be incorporated into the first protogalaxy. Indeed, other effects will become important in the course of the buildup
of the first galaxies. Among them is the ejection of metals into these systems by the first supernovae (e.g. Bromm et al. 2003).
However, we consider the early regime in which Pop III star formation dominates, and the effects of metals might not yet be important. Initially only taking into account the stellar radiative feedback, and neglecting chemical enrichment, relies on our simplifying assumption that only 100 {\rm M}$_{\odot}$ black hole-forming Pop III stars form
, which are predicted not to yield supernovae, and therefore not to eject metals into their surroundings (Heger et al. 2003).

Although the initial mass function (IMF) of the first generation of stars is not known with any certainty yet, there is mounting theoretical evidence that Pop III stars
were very massive, and thus it is very likely that many of these stars ended there lives by collapsing directly to black holes, emitting 
few or no metals into the IGM (see Fryer et al. 2001; Heger et al. 2003).  Here we assume that all of the stars that form within our cosmological box 
are black hole-forming stars which do not enrich the IGM with metals, and which therefore allow subsequent metal-free star formation to occur.  
Eventually, 
however, stars which create supernovae will form, and the ejected metals will be incorporated into the first protogalaxies, thus drawing the epoch of metal-free 
star formation to a close.  In light of this, we end our simulation after the formation of the eighth star in our box at a redshift of $z$ $\sim$ 
18, 
as we expect that at lower redshifts the effects of the first metals ejected into the primordial gas will become 
important (but see Jappsen et al. 2007).  Also, at lower redshifts global LW feedback, due to star formation at distances far larger than our cosmological box, will become increasingly 
important.  That said, by tracking the formation of individual Pop III stars in our box, we are able to find a variety of novel results concerning the 
local radiative feedback from the first generations of stars.     

\subsection{The First H~II Region and Lyman-Werner Bubble}
The first star appears in our cosmological box at a redshift of $z$ $\sim$ 23. It forms inside a minihalo with a total mass $\la$ 10$^6$ 
{\rm M}$_{\odot}$ and the gas within this halo is evaporated due to the photo-heating from the star. The H~II region that is 
formed around the star can be seen in Figure~2, which shows the electron fraction, H$_2$ fraction, temperature, and density of the gas, 
in projection.  The H~II region, which has a morphology similar to those found in previous studies, extends out to $\sim$ 4 kpc from the 
star, also similar to results found in previous works (Alvarez et al. 2006; Wise et al. 2006; Yoshida et al. 2007).

As shown in Figure~2, the molecules within $\sim$ 5 kpc are photodissociated by the LW radiation from the first 
star, and the LW bubble extends to only $\sim$ 1 kpc outside of the H~II region.  Noting that the formation timescale for H$_2$ in the neutral IGM 
at these redshifts is of the order of $\sim$ 300 
Myr, much longer than the lifetime of the massive stars that we consider here, we can estimate the distance through the IGM that 
the LW bubble should extend, $R_{\rm LW}$, by evaluating the criterion for the effective dissociation of molecules at this distance from the first star: $t_{\rm diss, H_2}$ = $t_{\rm *}$.  
Using equation (8) and taking the lifetime of the star to be 3 Myr gives $R_{\rm LW}$ $\sim$ 5 kpc, consistent with the result we find for the first star, shown in Figure~2 (see also Ferrara 1998). 
Outside of this LW bubble molecules will not be dissociated effectively by the single Pop III star, owing largely to its short lifetime.  Only when continuous star formation sets in will the LW bubbles of the first generations of stars merge and become large enough to establish a more pervasive LW background flux (see e.g. Haiman et al. 2000).  

\subsection{Thermal and Chemical Evolution of the Gas}
The properties of the primordial gas within our box are strongly time-dependent, as any gradual evolution of the gas is disrupted each time 
a star turns on and heats the gas, ionizes atoms, and photodissociates molecules.  Certain robust patterns, however, do emerge in the 
course of the evolution of the primordial gas. Figure~3 shows the chemical and thermal properties of the gas at a redshift $z$ $\sim$ 18, 
just after the death of the eighth star in our cosmological box. Here, the light-shaded particles are those which have been 
contained within an H~II region, and so have passed through a fully ionized phase.  

The ionized gas in the H~II regions begins to recombine and cool once the central star dies. The dynamical expansion of these hot 
regions leads to the adiabatic cooling of the gas, as can be seen in the upper left panel of Figure~3. The plot shows relic H~II regions at different evolutionary
stages. The older ones are generally cooler, owing to the molecular cooling that has had 
more time to lower the temperature of the gas. Indeed, the first relic H~II regions by this redshift, $\sim 70$~Myr 
after the first star formed, have already cooled to near the temperature of the un-ionized gas.  The electron fraction of the relic H~II region 
gas, however, is still much higher than that of the un-ionized gas, as can be seen in the upper-right panel of Figure~3. That the 
cooling of the gas occurs faster than its recombination leads to the rapid formation of molecules (e.g. Kang \& Shapiro 1992; 
Oh \& Haiman 2003; Nagakura \& Omukai 2005; Johnson \& Bromm 2006). This elevated fraction of both H$_2$ and HD molecules in the 
relic H~II region gas is evident in the bottom panels of Figure~3.

The high abundance of molecules in relic H~II regions can lead to efficient cooling of the gas, and this has important consequences 
in the first protogalaxies. In particular, a high fraction of HD in these regions could allow the gas to cool to the temperature of 
the cosmic microwave background (CMB), $T_{\rm CMB}$, the lowest temperature attainable by radiative cooling, and this effective cooling may lead to the 
formation of lower mass metal-free stars (Nagakura \& Omukai 2005; Johnson \& Bromm 2006; Yoshida 2006). Indeed, Figure~3 shows that 
the HD fraction can greatly exceed the minimum value needed for efficient cooling to the CMB temperature floor in local thermodynamic 
equilibrium (LTE), $f_{\rm HD, crit}$ $\sim$ 10$^{-8}$ (Johnson \& Bromm 2006).         

While the LW feedback from the stars that form in our box can very effectively destroy molecules within $\sim$ 5 kpc of the stars by the end of their lives, this feedback is not 
continuous. Following the death of a given star, the molecules will begin to reform in the absence of LW radiation.
The time required for the formation of H$_2$ molecules is sensitively dependent on the ionized fraction of the gas, but the formation 
time can be relatively short for un-ionized gas at high densities, as well.  In relic H~II regions the fraction of H$_2$ can reach 
10$^{-4}$ within $\sim$ 1 Myr (Johnson \& Bromm 2007). 
In collapsing minihalos, where the molecules play a key role in cooling the 
gas and allowing it to continue collapsing, the formation times are in general longer at the densities we consider here, $n$ $\la$ 20 cm$^{-3}$.  We find 
that the formation timescale for un-ionized gas collapsing in minihalos is $t_{\rm form, H_2}$ $\sim$ 5 $\times$ 10$^5$ yr 
at a density of 1 cm$^{-3}$ and a temperature of 
900 K, and $t_{\rm form, H_2}$ $\sim$ 7 $\times$ 10$^6$ yr at a density of 0.1 cm$^{-3}$ and a temperature of 500 K.
The average time between the formation of stars in our box is $\sim 10$~Myr, and so the molecules inside sufficiently dense minihalos 
can often reform and allow the gas to continue cooling and collapsing, in spite of the intermittent LW feedback from local star 
forming regions.  

In order to evaluate the possible effects of continuous LW feedback from sources outside of our box, we have carried out simulations in which we include a LW background
which destroys H$_2$ molecules at a rate given by (Abel et al. 1997)

\begin{equation}
k_{\rm diss} = 1.2 \times 10^{-12} J_{\rm LW}  {\rm \,s}^{-1}\mbox{\ ,}
\end{equation} 
where $J_{\rm LW}$ is the flux of LW photons in units of 10$^{-21}$ ergs s$^{-1}$ cm$^{-2}$ Hz$^{-1}$ sr$^{-1}$.  We have carried out simulations in which the value of $J_{\rm LW}$ is taken to be zero before the formation of the first star and 0.1, 10$^{-2}$, and 10$^{-3}$ afterwards, when a LW background might be expected to begin building up due to distant star formation.  For each of these simulations, we found the formation redshift of the second star in the box to be $z_{\rm 2nd}$ = 16.3, 20, and 20.5, respectively.  In our main simulation, in which we neglect a possible background LW flux, the second star formed at $z_{\rm 2nd}$ = 20.6.  This demonstrates that a background LW flux of $J_{\rm LW}$ $\la$ 10$^{-2}$ would likely have little impact on our results, while a larger LW flux would simply delay the collapse of gas into minihalos and so lower the overall star formation rate in our box, consistent with previous findings (see e.g. Machacek et al. 2001; Mesinger et al. 2006).  We emphasize, however, that for a substantial LW background to be established, a relatively high continuous star formation rate must be achieved, as we have shown that individual Pop III stars can only be expected to destroy molecules within $\sim$ 5 kpc of their formation sites.  In the very early stages of the first star formation, when short-lived single stars are forming in individual minihalos (see e.g. Yoshida et al. 2006), it appears unlikely that a substantial LW background would be established, the feedback from the sources being instead largely local.  It may be only later, when continuous star formation begins to occur in larger mass systems that a pervasive LW background would likely be built up (see, e.g., Haiman et al. 2000; Greif \& Bromm 2006).      


\subsection{Shielding of Molecules by Relic H~II Regions}
As star formation continues, the volume occupied by relic H~II regions increases. Because of the high molecule fraction that can 
develop in these regions, owing to the large electron fraction that persists for $\la$ 500 Myr (Johnson \& Bromm 2007), 
the increasing volume of the IGM occupied by relic H~II regions implies an increase in the optical depth to LW photons in the vicinity of the first star formation sites.
By a redshift of $z$ $\sim$ 18, eight stars have formed in our cosmological box and each has left behind a relic H~II region.   

 As can be seen in Figures~3 and 4, the gas 
inside the relic H~II regions that have formed contains an H$_2$ fraction generally higher than the primordial 
abundance of 10$^{-6}$, and up to an abundance of $\sim$ 10$^{-3}$ in the denser regions.  This elevated fraction of H$_2$ inside 
the relic H~II regions leads to a high optical depth to LW photons, $\tau_{\rm LW}$, through the relic H~II regions.  The column density 
through a relic H~II region which recombines in the absence of LW radiation can become of the order of $N_{\rm H_2} \sim$ 10$^{15}$ cm$^{-2}$
(Johnson \& Bromm 2007).  Because the molecules in the relic H~II regions that we consider here are subject to LW 
feedback from neighboring star formation regions, the optical depth through these regions may in general be lower.  However, 
the rapid rate of molecule formation in these regions, even considering the LW feedback from local star forming regions in our 
box, allows the molecule fraction to approach 10$^{-4}$ as late as $\sim$ 100 Myr after the 
death of the central star.  This elevated molecule fraction combined with the growing volume-filling fraction of relic H~II regions 
leads to an appreciable optical depth to LW photons, which generally increases with time as more stars form and create more relic 
H~II regions.  
To quantify this effect, we calculate the average column density of H$_2$ molecules through a cubic region of side length $l$ as
the product of the length $l$ and the volume averaged number density of H$_2$ molecules, given by

\begin{equation}
N_{\rm H_2} \simeq l \frac{\sum n_{\rm H_2} V}{\sum V}  \mbox{\ ,}
\end{equation}
where the sum is over all of the SPH particles in the volume and $n_{\rm H_2}$ is the number density of H$_2$ at each of the SPH particles.  The volume associated with each individual 
SPH particle, $V$, is estimated as $V$ $\simeq$ m$_{\rm SPH}$/$\rho$, where m$_{\rm SPH}$ is the mass of the SPH particle and $\rho$ is the mass density of the 
gas at that particle.  The optical depth to LW photons is then computed as (Draine \& Bertoldi 1996; Haiman et al. 2000)

\begin{equation}
\tau_{\rm LW} \simeq 0.75 \ln(\frac{N_{\rm H_2}}{10^{14} {\rm cm}^{-2}})  \mbox{\ .}
\end{equation}

Figure~5 shows the optical depth to LW photons avergaged both over the central comoving 153 kpc $h^{-1}$ of our cosmological box, in which the first star forms, and over the entire box, for which the comoving side length is 460 kpc $h^{-1}$ .  Before the formation of the first star, the optical depth evolves largely owing to the cosmic expansion, following the relation $\tau_{\rm LW}$ $\simeq$ $n_{\rm H_2}$ $l$ $\propto$ (1+$z$)$^2$, because the average H$_2$ fraction does not change appreciably.  However, with the formation of the first star in our box at $z$ $\simeq$ 23 the optical depth begins to change dramatically in the inner portion of the box, first falling to a value of $\simeq$ 0.1 due to the LW feedback from the first star and then steadily climbing to values $\ga$ 2 as copious amounts of molecules form inside the relic H~II regions that accumulate as star formation continues.  

The evolution of the optical depth averaged over the entire box is not as dramatic, as the fraction of the volume of the whole box occupied by relic H~II regions is much smaller than the fraction of the central region that is occupied by these molecule-rich regions.  However, the optical depth averaged over the whole box, which is a better estimate of the optical depth over cosmological distances, still rises to $\tau_{\rm LW}$ $\ga$ 1.5 across our box, an appreciable value which will serve to impede the build-up of a cosmologically pervasive LW background.       


\subsection{Black hole accretion}
Accretion onto Pop III relic black holes may be inefficient for some time following the formation of these objects, owing to the fact that Pop III 
stars photo-heat and evaporate the gas within the minihalos which host them (Johnson \& Bromm 2007; see also Yoshida 2006).
Indeed, accretion onto Pop III relic black holes at close to the Eddington limit can only occur if the accreted gas has a density 
above $\sim$ 10$^2$ cm$^{-3}$, and it is only in collapsing halos that such densities are achieved at the high redshifts at which the 
first stars formed (Johnson \& Bromm 2007).  By assumption, all of the stars that are formed in our simulation are black hole-forming 
Pop III stars.  If these black holes remain inside their host minihalos, then by tracking the evolution of the gas within these 
photo-evaporated host minihalos, we can learn when efficient accretion onto these Pop III relic black holes may occur.  

The minihalo within which the first star forms at a redshift $z$ $\sim$ 23 resides within the relic H~II region left by the first star.  
Due to the formation of a high fraction of molecules, and to the molecular cooling that ensues, the relic H~II region gas cools down to 
temperatures $\sim$ 10$^3$ K, below the virial temperature of this 10$^6$ {\rm M}$_{\odot}$ minihalo.  The gas then re-collapses into the 
minihalo, reaching a peak density of $n_{\rm res}$ at a redshift $z$ $\sim$ 19, or $\sim$ 50 Myr after the formation of the first star.  
Figure~6 shows the properties of the relic H~II region gas as a function of distance from the center of this minihalo, at the time when the 
gas has collapsed to a density of $n_{\rm res}$.  We cannot, with this simulation, resolve what happens once the gas collapses further and 
reaches higher densities.  However, we can estimate the time it will take for the gas at the center of the halo to reach a density of $n$ 
$\sim$ 10$^2$ cm$^{-3}$ as the free fall time of the gas, which is $t_{\rm ff}$ $\sim$ 10 Myr.  Thus, a Pop III relic black hole at the 
center of this halo could be expected to begin accreting gas efficiently $\sim$ 60 Myr after its formation.  This is a significant delay, 
and could pose serious challenges to theories which predict that efficient accretion onto Pop III relic black holes can lead to these 
black holes becoming the supermassive black holes that power the quasars observed in the {\it Sloan Digital Sky Survey} at redshifts 
$z$ $\ga$ 6 (e.g. Yoshida 2006; Johnson \& Bromm 2007; Li et al. 2006).

\subsection{HD cooling in relic H~II regions}
While abundant molecules can form within relic H~II regions, the LW feedback from neighboring star-forming regions can suppress 
the effect of this elevated fraction of molecules.  The electron fraction remains high in relic H~II regions for up to $\sim$ 500 Myr in 
the general IGM, but in higher density regions where the gas is recollapsing,  the electron fraction drops much more quickly. Figures~3 and 6 
show that the electron fraction drops to a value of $\la$ 10$^{-4}$, comparable to the electron fraction of the un-ionized gas, once the density 
of the relic H~II region gas becomes 
$\ga$ 10 cm$^{-3}$.  Thus, once the gas reaches these densities the ionized fraction will become too low to catalyze the formation of a high 
fraction of molecules, and of HD molecules in particular. Therefore, in order for HD to be an effective coolant of the primordial gas in relic H~II 
regions, the abundanct HD molecules that are formed at densities $\la$ 10 cm$^{-3}$ must not be destroyed by LW feedback from neighboring 
star-forming regions before the gas collapses to high densities and forms stars.  If we estimate the timescale on which the relic H~II region gas would 
collapse to form stars as the free-fall time of the gas, we find that the molecules must be shielded from photodissociating radiation for at least $t_{\rm ff}$ 
$\ga$ 10 Myr in order for the high abundance of HD molecules to persist, so that the formation of so-called Pop~II.5 stars might be enabled, with their hypothesized masses of $\sim 10 {\rm \,M}_{\odot}$ (see Johnson \& Bromm 2006).

We find that the relic H~II region gas that re-collapses into the minihalo in which the first star formed, shown in Figure~6, carries 
a high fraction of HD molecules, as LW feedback from neighboring stars does not effectively dissociate the molecules in this relatively dense and self-shielded gas.  The HD fraction exceeds 10$^{-7}$, becoming an order of magnitude higher than its value for un-ionized primordial gas collapsing in a minihalo. Thus, 
for this case, HD cooling will likely be effective at higher densities as the gas collapses further, and we expect that a Pop II.5 star, 
with a mass of the order of 10 {\rm M}$_{\odot}$, might form later on, if we were to run the simulation further (Nagakura \& Omukai 2005; Yoshida 2006; Johnson \& Bromm 2006; Yoshida et al. 2007).   Had star formation taken place nearer this minihalo between the formation of the first star and the 
re-collapse of the gas into the host minihalo, then the molecule fraction would likely not be so elevated, and a higher mass metal-free star would be
more likely to form.  Thus, while in our simulation it appears that the first relic H~II region that forms may give rise to Pop II.5 star formation, we emphasize that the possibility of the formation of Pop II.5 stars in relic H~II regions is very dependent on the specific LW feedback that affects the gas in these regions.     

\subsection{Star Formation in the Presence of Radiative Feedback}
To discern the effect the local radiative feedback from the first stars has on the star formation rate, we have compared the results 
obtained from our simulations with and without radiative feedback.  By a redshift of $z$ $\sim$ 18, a total of nine 
star-forming regions were identified in our simulation without feedback, while at the same epoch eight stars had formed in our 
simulation including feedback.  Thus, we find that the average star formation rate at redshifts $z$ $\ga$ 18 is diminished by a 
factor of perhaps only $\la$ 20 percent due to local radiative feedback, although this result is subject to the small number statistics 
within our single cosmological box. 
Figure~7 shows the locations of the sites of star formation for both cases, 
plotted in comoving coordinates against the projected density field.  The orange squares denote sites where Pop III stars could have 
formed in the case without feedback, while the green dots denote sites where Pop III stars formed in the simulation including radiative 
feedback.  Thus, the sites where star formation is suppressed by the radiative feedback are marked by the orange squares which are 
not filled by a green dot.


We point out, however, that we do not include LW feedback from stars which may have formed outside of our box, and hence it is possible that the overall LW feedback may 
be stronger than we find here.  At redshift $z$ $\sim$ 18, we end the simulation, but note 
that star formation will likely take place at an increasing rate as the collapse fraction increases with time.  This could lead to a continuous LW 
background produced within our box, different from the intermittent LW feedback produced by individual stars that occurs in the simulation down 
to $z$ $\sim$ 18.  

Also, the limits of our resolution 
prohibit us from discerning the stronger shielding of H$_2$ molecules and I-front trapping that could occur within very dense 
collapsing minihalos (see Ahn \& Shapiro 2006). 
However, we note that these authors find that the radiative feedback on collapsing minihalos from
nearby stars generally does not greatly affect the final outcome of the collapse, as halos which collapse in the absence of radiative feedback generally
also collapse when radiative feedback is applied (see also Susa \& Umemura 2006), roughly consistent with our results in the present work. 
Thus, our limited resolution may not substantially impact the results that we find for the slight suppression of star formation due to local radiative feedback, 
although higher resolution simulations will 
be necessary to more precisely study the full impact of radiative feedback from the first stars on the first protogalaxies.    

\section{Summary and Discussion}
We have performed cosmological simulations which self-consistently account for the radiative feedback from individual Pop III stars, as they form 
in the course of 
the assembly of the first protogalaxies. 
We have solved in detail for the H~II regions, as well as for the LW bubbles of 
these stars, wherein molecule-dissociating radiation effectively destroys H$_2$ and HD molecules.  The local radiative feedback 
from the first stars is complex, and we find a variety of novel results on the evolution of the primordial gas, on the effects of the LW 
radiation from the first stars, on the nature of second generation star formation, and on black hole accretion.

While the LW radiation from the first stars can, in principle, greatly suppress Pop III star formation in the early universe, we find that a 
number of factors minimize the effectiveness of this negative feedback.  
Firstly, the LW radiation produced locally by individual 
stars is not uniform and constant, as LW feedback has been modeled in previous work (e.g., Ciardi \& Ferrara 2005; Mesinger et al. 2006), but rather is present only during 
the brief lifetimes of the individual stars that produce it. Thus, even if the molecules in collapsing minihalos and relic H~II regions are 
destroyed by the radiation from individual stars, they will, at the early stages of Pop III star formation, have time to reform and continue 
cooling the primordial gas in between the times of formation of local stars.  Furthermore, because the LW bubbles of individual Pop III stars extend only to
$R_{\rm LW}$ $\sim$ 5 kpc from these sources, due to the short stellar lifetimes, the build up of a pervasive LW background would likely have to await the epoch of continuous star formation, which is fundamentally different from the epoch of the first stars, in which these sources shine for short periods within individual minihalos.   

As star formation continues, the volume-filling 
fraction of relic H~II regions increases as well, and this, combined with the high fraction of molecules that form in these regions, leads to an 
opacity to LW photons through the IGM which increases with time.  This opacity can become of the order of $\tau_{\rm LW}$ $\ga$ 2 through 
individual relic H~II regions (Johnson \& Bromm 2007; see also Ricotti et al. 2001; Machacek et al. 2001, 2003; Oh \& Haiman 2002).  Furthermore, as the 
volume-filling fraction of relic H~II regions increases with time, $\tau_{\rm LW}$ through the general IGM may become similarly large, and this effect 
will have to be considered in future work which seeks to elucidate the effect of LW feedback on Pop III star formation. 

We find that metal-free stars with masses of the order of 10 {\rm M}$_{\odot}$, the postulated Pop~II.5 stars 
(e.g. Johnson \& Bromm 2006; Greif \& Bromm 2006), might form from the molecule-enriched gas within the first relic H~II regions, although 
we note that this may not occur in general, due to LW feedback from neighboring star-forming regions.  This susceptibility to LW feedback is due to the fact that the high fraction of HD molecules which forms in the electron-rich, low density regions of relic H~II regions must persist until this gas has had time to collapse to high densities and form stars.  If LW feedback from neighboring star-forming regions destroys the molecules after the gas has collapsed to densities $\ga$ 10 cm$^{-3}$, then the abundance of HD molecules will not likely be elevated when the gas forms stars and Pop II.5 star formation may be suppressed.  Because this implies that the molecules must be shielded from LW radiation for, at least, the free-fall time for gas with densities $\la$ 10 cm$^{-3}$, or $\ga$ 10 Myr, we conclude that Pop II.5 star formation in relic H~II regions may occur only in the circumstances when local Pop III star formation is suppressed over such timescales.  However, we also point out that the shielding provided by the high H$_2$ fraction in these relic H~II regions may help to minimize the LW feedback from neighboring star-forming regions, and so may make Pop II.5 star formation in relic H~II regions possible in many cases. 

We find that the ionized primordial gas surrounding the first star formed in our simulation, at a redshift of $z$ $\sim$ 23, recombines and cools by molecular cooling to temperatures below the virial temperature of the minihalo that hosted this first star. Thus, this relic H~II region gas is able to re-collapse to densities $\ga$ 20 cm$^{-3}$ within this minihalo after $\sim$ 50 Myr from the death of the star.  It is predicted that many Pop III stars will collapse directly to form black holes with masses of the order of 100 {\rm M}$_{\odot}$ (e.g. Heger et al. 2003), and if such a black hole resides within this host halo, then we find that it may begin accreting dense primordial gas at close to the Eddington rate after $\sim$ 60 Myr from the time of its formation.  This is an important consideration to be incorporated into models of the growth of the $10^9 {\rm \, M}_{\odot}$ black holes which have been observed at redshifts $\ga 6$, as it places constraints on the amount of matter that a given relic Pop III black hole could accrete by this redshift (e.g. Haiman \& Loeb 2001; Volonteri \& Rees 2006; Li et al. 2006).

Finally, by comparing the star formation rates which we derive from our simulation including radiative feedback with those derived from our simulation in which feedback is left out, we have seen that local radiative feedback from the first stars likely only diminishes the Pop III star formation rate by a factor of, at most, a few.  In our simulation, in particular, we find that this rate is decreased by only $\la$ 20 percent, although this may be less suppression than would be expected by the overall radiative feedback, as we did not include the possible effects of a global LW background.  Future simulations which resolve densities higher than those reached here, and which self-consistently track the build-up of the LW background along with the IGM opacity to LW radiation, will be necessary to more fully explore the radiative effects of the first stars on the formation of the first galaxies.    
However, the goal of understanding the formation of the first galaxies is now
clearly getting within reach, and the pace of progress is expected to be rapid. 

\acknowledgments{
We would like to thank Yuexing Li and Kyungjin Ahn for helpful discussions.  We are also grateful to Simon Glover and the anonymous referee for valuable comments that improved the quality of this work.  The simulations used in this work were carried out at the Texas Advanced 
Computing Center (TACC).
}



\clearpage

\begin{figure}
\plotone{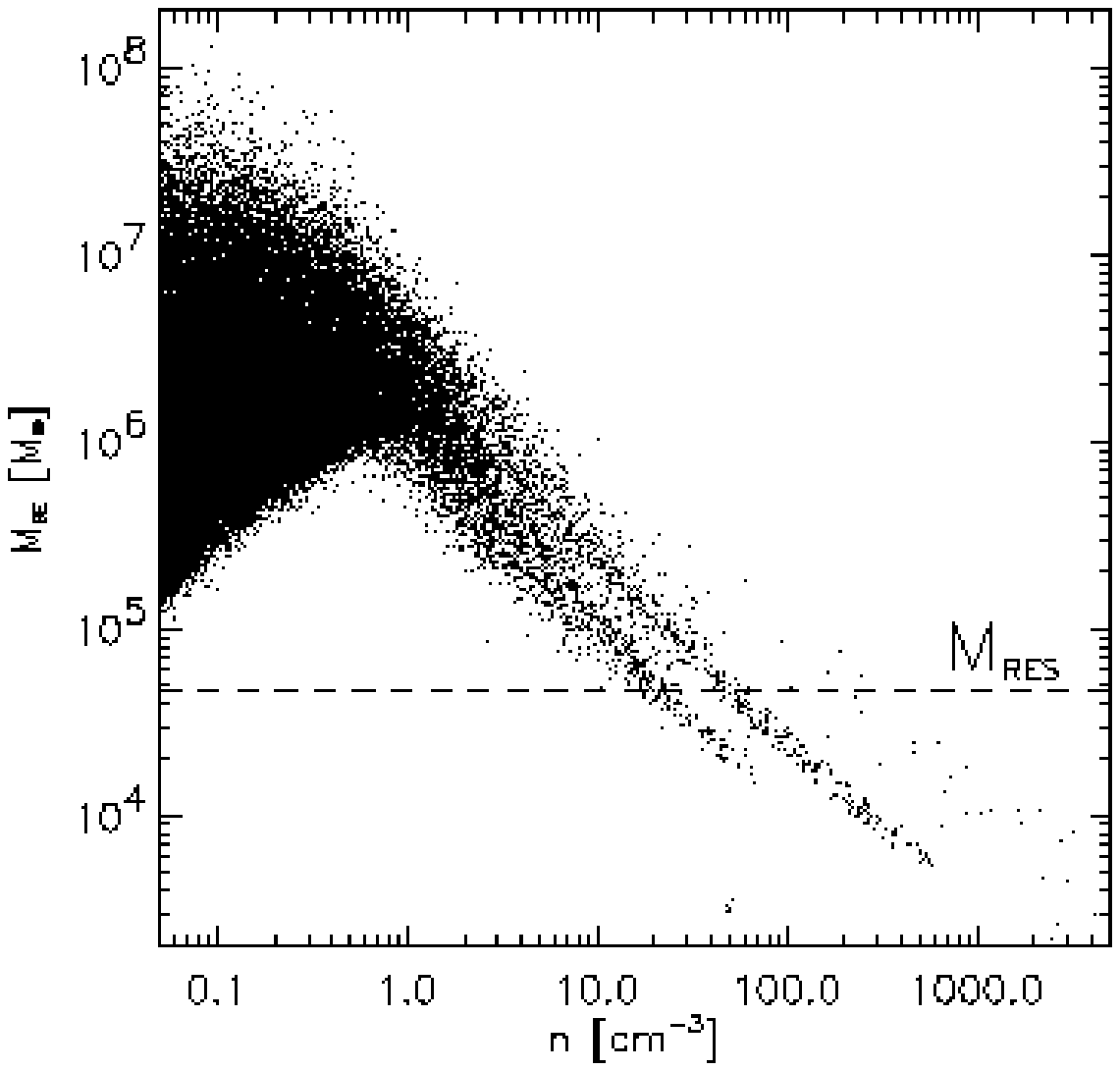}
\caption{Determining the maximum density resolvable in our simulations.  To reliably resolve the properties of the gas in our simulation, the Bonnor-Ebert mass, 
similar to the Jeans mass, must be larger than the mass in the SPH smoothing kernel.  For added assurance, we take the minimum resolvable mass to be 
twice the mass in the kernel.  This value for the resolution mass is shown by the dashed horizontal line.  For densities higher than $n_{\rm res}$ $\sim$ 
20 cm$^{-3}$, the Bonnor-Ebert mass may be exceeded by the resolution mass, and so we take it that we can only resolve the properties of the gas at densities 
below this value.  We note that the two structures emerging at high densities are two spatially distinct halos of different mass which are undergoing collapse.
\label{fig1}}
\end{figure}

\begin{figure}
\plotone{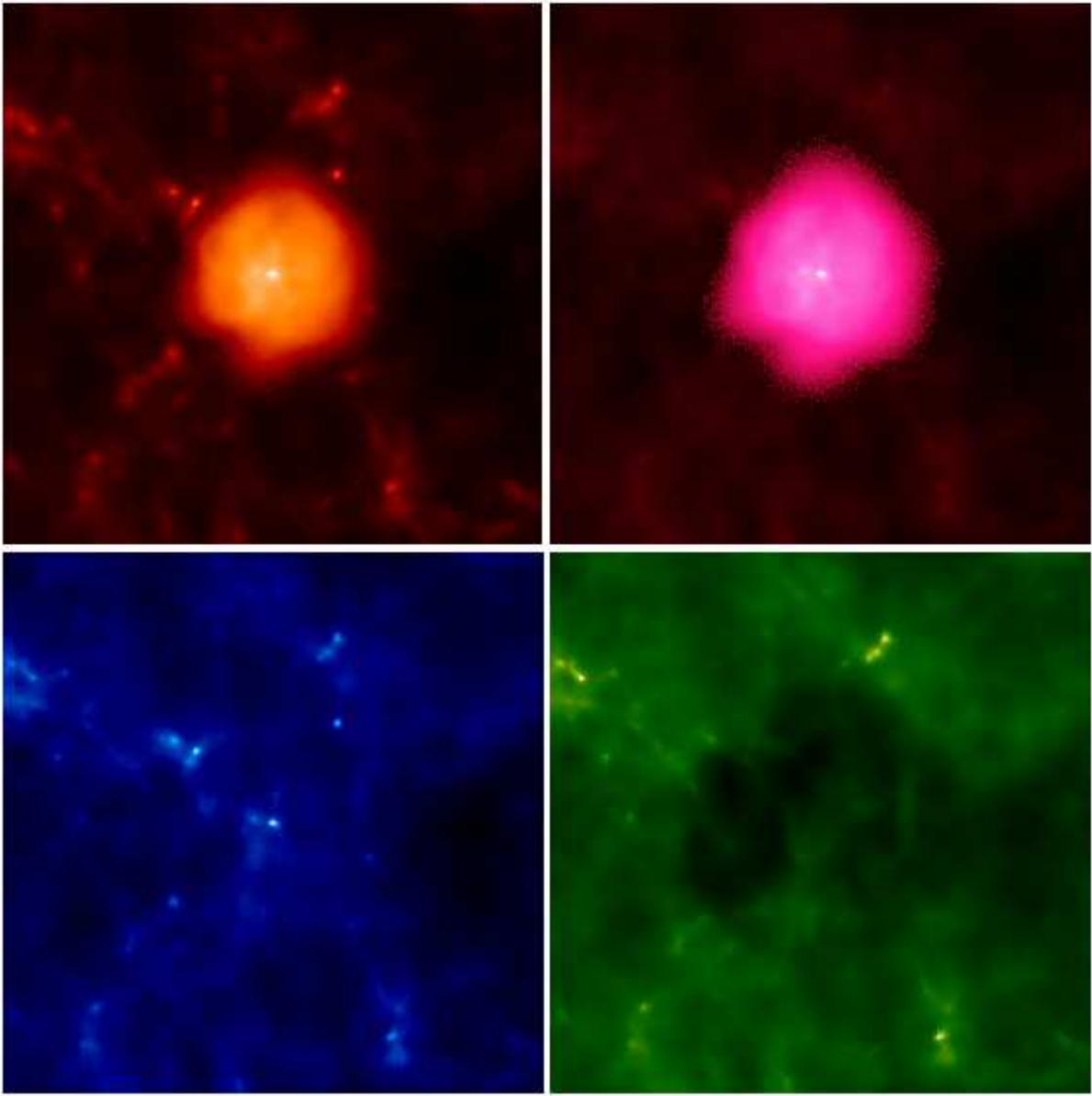}
\caption{The first H~II region and LW bubble.  Clockwise from top-left are the temperature, electron fraction, H$_2$ fraction, and density, 
plotted in projection.  While the size of our cosmological box is $\sim$ 27 kpc in physical units at this redshift, $z$ $\sim$ 
23, here we have zoomed into the inner 20 kpc, in order to see detail around the first star.  The H~II region extends out to $\sim$ 4 kpc in 
radius, while the LW bubble extends to $\sim$ 5 kpc, within which the molecule fraction is zero.  In each panel, the lighter 
shades signify higher values of the quantity plotted.     
\label{fig2}}
\end{figure} 

\begin{figure}
\plotone{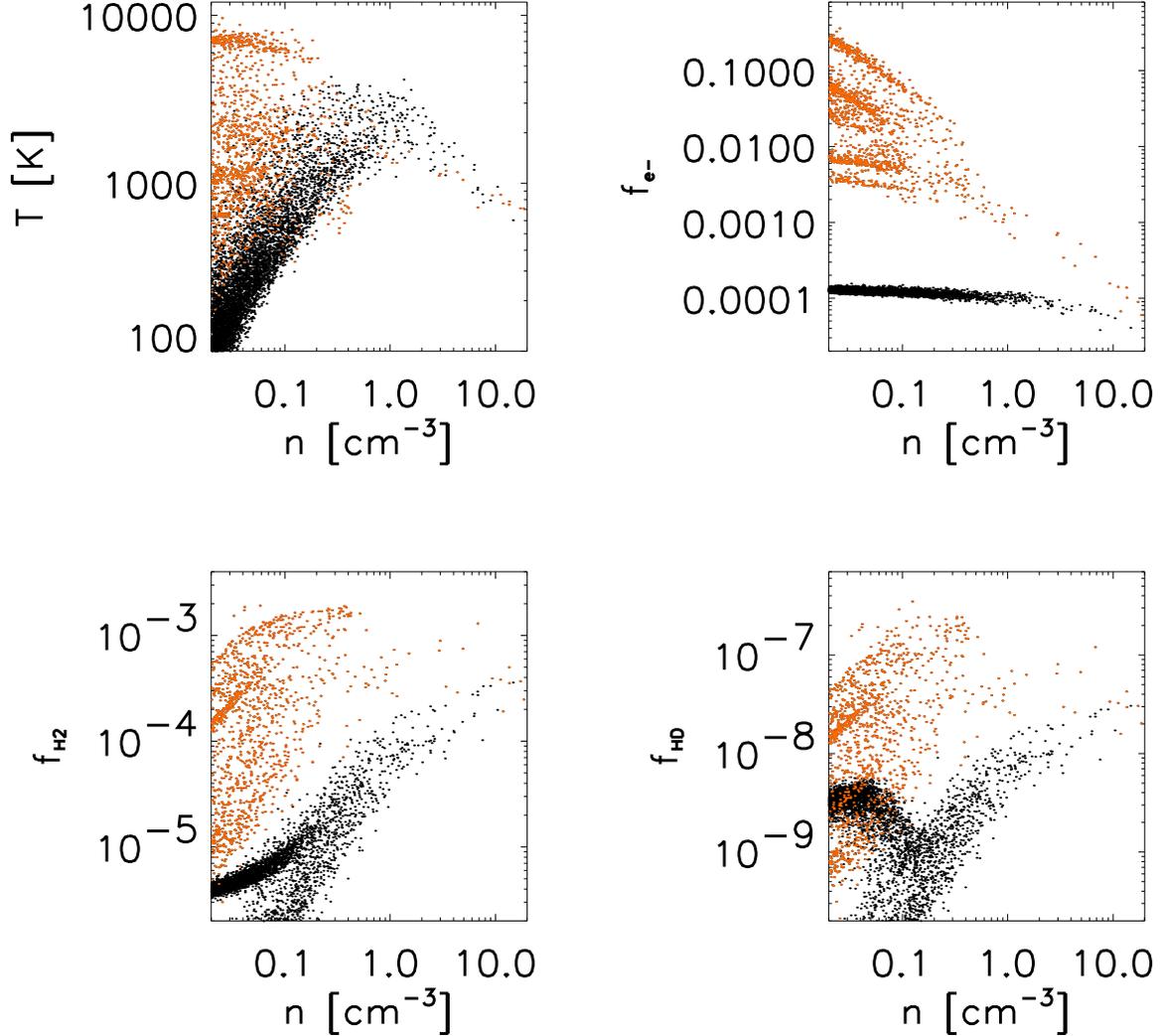}
\caption{The properties of the primordial gas at redshift $z$ $\sim$ 18, at the end of the life of the eighth star. The SPH particles which have experienced an ionized phase 
within an H~II region are colored in orange (gray), while those that have not are in black. Clockwise from the top-left, the temperature, 
free electron fraction, HD fraction, and H$_2$ fraction are plotted as functions of gas density. The relic H~II region gas 
cools largely by adiabatic expansion, but, importantly, also by cooling facilitated by the high abundance of H$_2$ and HD molecules, 
which arises owing to the high electron fraction in this gas.  The high electron fraction persists until the gas has collapsed to 
densities of $\ga$ 10 cm$^{-3}$, as can be seen in the top-right panel.  The molecule fraction is highest at low densities for gas 
in which the molecules have not been destroyed by LW feedback, giving rise to the features seen at low densities in the bottom two panels.  
\label{fig3}}
\end{figure} 

\begin{figure}
\plotone{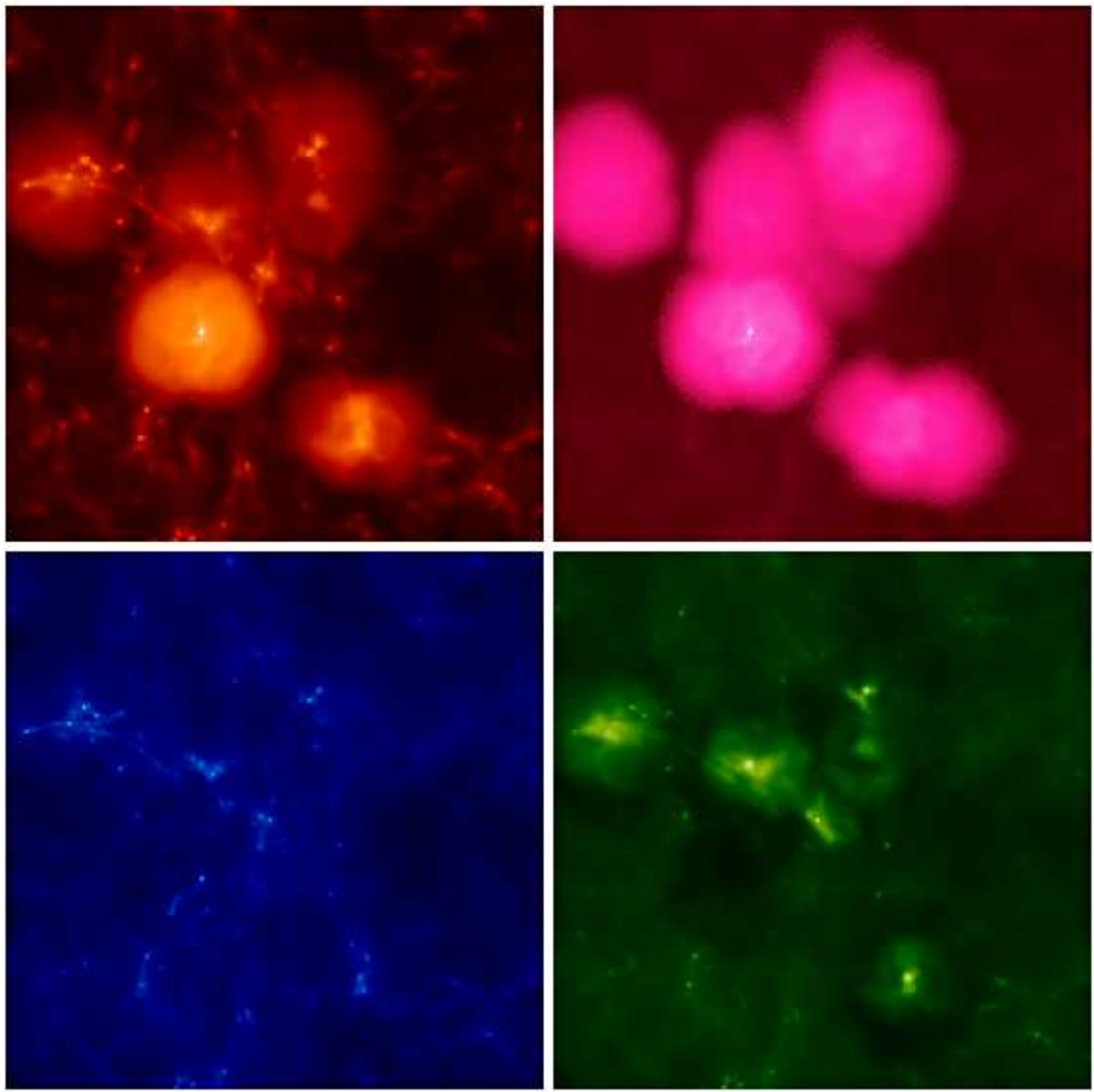}
\caption{The properties of the primordial gas at the end of the life of the eighth star in our box, at redshift $z$ $\sim$ 18. Clockwise from top-left are the temperature, electron fraction, H$_2$ fraction, and density of the gas, in projection. Here we show the entire cosmological box, which is $\sim$ 35 kpc in physical units. Note the high electron and H$_2$ fractions in the relic H~II regions, where recombination is taking place. The elevated H$_2$ fraction in these regions raises the optical depth to LW photons through them significantly, as is illustrated in  Fig. 5.  The temperature in the older relic H~II regions is not greatly elevated as compared to the temperature of the un-ionized gas, owing to the adiabatic and molecular cooling that takes place in these regions.  The lighter shades denote higher values of the quantities plotted.               
\label{fig4}}
\end{figure} 

\begin{figure}
\plotone{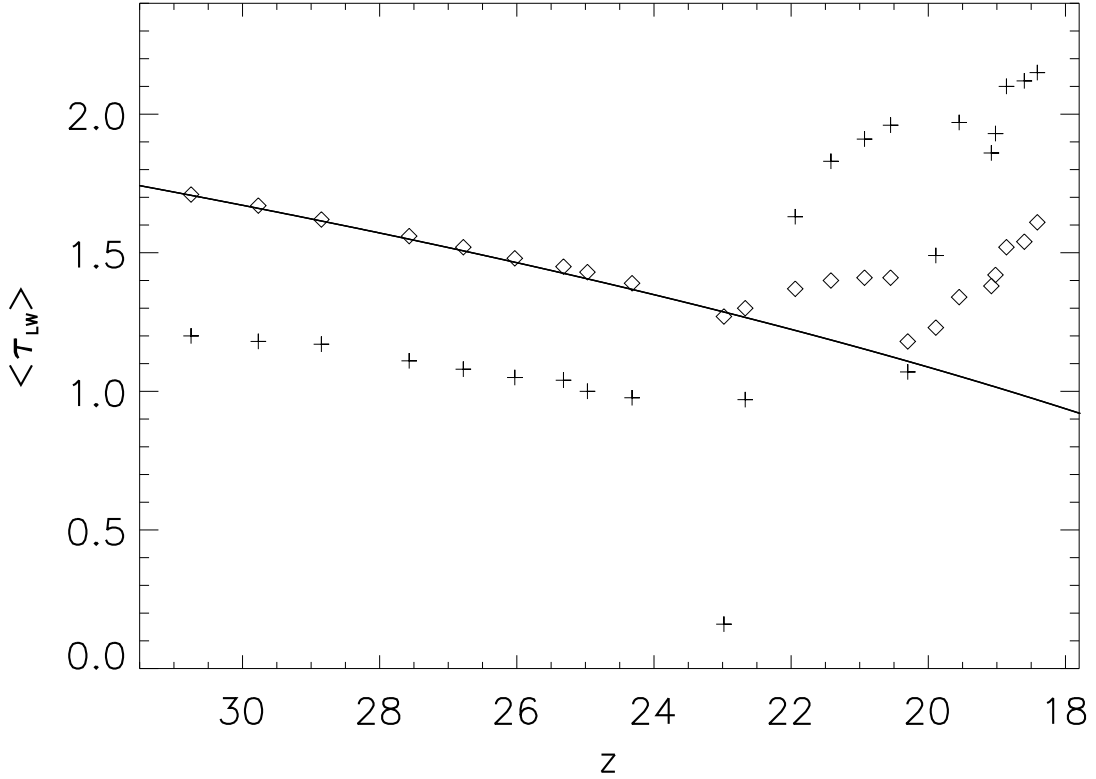}
\caption{The optical depth to LW photons, $\tau_{\rm LW}$, averaged over two volumes in our box, as a function of redshift, $z$.  The diamonds denote the optical depth averaged over the entire cosmological box, while the crosses denote the optical depth averaged only over a cube containing the inner comoving 153 kpc $h^{-1}$ of the box, centered in the middle of the box with a volume one ninth that of the whole box.  It is within this region that the first star forms and the star formation rate is higher than the average star formation rate over the whole box, and this is reflected in the higher local optical depth in this region as relic H~II regions accumulate in the box.  The average optical depth through the entire box also rises, but the increase is less dramatic.  The solid line denotes the optical depth to LW photons, averaged over the whole box, that would be expected for the case that the gas maintains the average cosmological density everywhere and that the H$_2$ fraction does not change from the primordial value of 2 $\times$ 10$^{-6}$; for this case, the optical depth changes owing only to cosmic expansion.  Note that the optical depth averaged over the whole box matches well this idealized case up until the first star forms at a redshift of $z$ $\sim$ 23.  The temporary drops in the optical depth occur due to LW feedback when individual stars form.  
\label{fig5}}
\end{figure}

\begin{figure}
\plotone{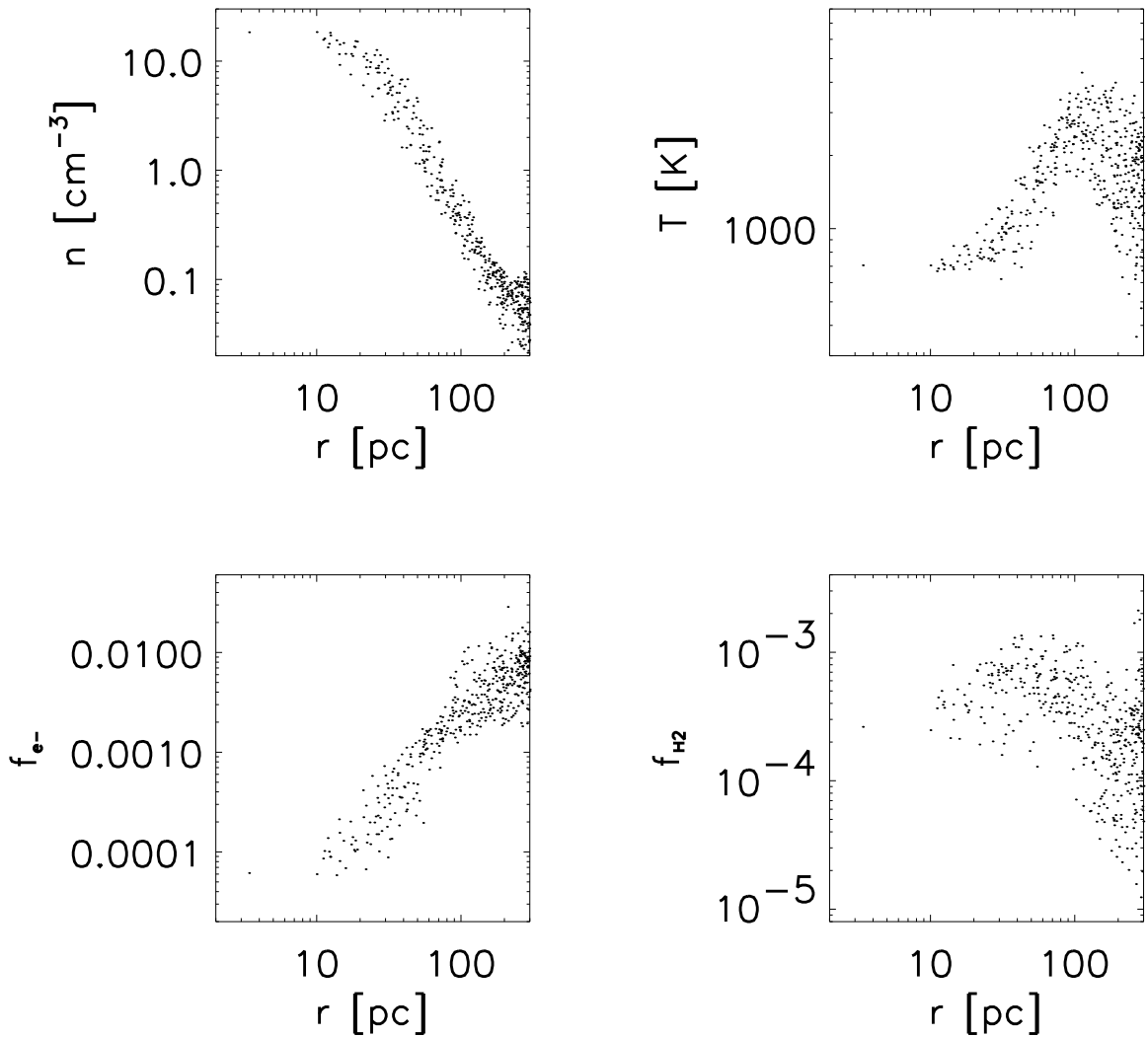}
\caption{The properties of the relic H~II region gas which recollapses into the minihalo which hosted the first star. Clockwise from 
the top-left are the density, temperature, H$_2$ fraction, and free electron fraction plotted as functions of distance from the center of the 
minihalo at the time when the density reaches $n_{\rm res}$, at a redshift of $z$ $\sim$ 19. The temperature of the gas has dropped to 
below 10$^3$ K, well below the 
virial temperature of the minihalo, owing to molecular cooling.  Owing to the high electron fraction that persists in this relic H~II region, 
the molecule fraction in this gas is higher than in the case of un-ionized primordial gas 
collapsing into a minihalo.  Indeed, as can be seen in Figure~3, the HD fraction is roughly an order of magnitude higher at these densities than in the case of un-ionized gas collapsing 
into a minihalo, which may allow for the efficient cooling of the gas to temperatures $T$ $\ga$ $T_{\rm CMB}$ and so perhaps for the formation of metal-free stars with masses of the order of 10 {\rm M}$_{\odot}$.  
\label{fig6}}
\end{figure}

\begin{figure}
\plotone{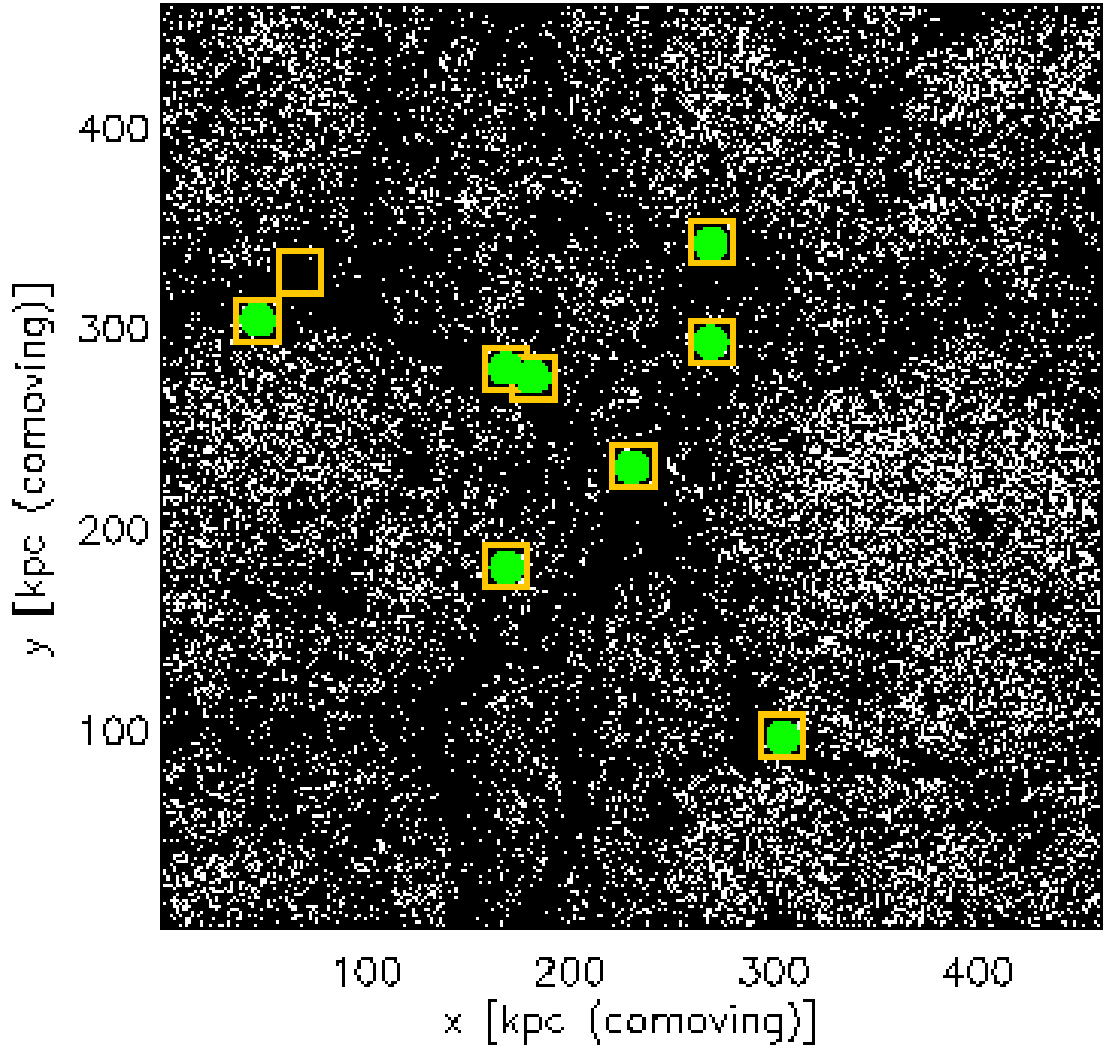}
\caption{The sites of star formation with and without radiative feedback, at redshift $z$ $\sim$ 18.  The black dots show the density field in our simulation box, 
in projection. The orange squares show the locations of minihalos in which Pop III star formation could take place, in our simulation 
without radiative feedback.  The green dots show the sites where star formation takes place in our simulation including radiative 
feedback. 
\label{fig7}}
\end{figure}


\end{document}